\documentclass{ws-ijmpd}
\usepackage[super,compress]{cite}

\newcommand{\be}{\begin{equation}}
\newcommand{\ee}{\end{equation}}

\begin{document}
\markboth{G.S. Bisnovatyi-Kogan, M. Merafina}
{Two body problem in presence of cosmological constant}

\catchline{}{}{}{}{}
%
%%%%%%%%%%%%%%%%%%%%%%%%%%%%%%%%%%%%%%%%%%%%%%%%%%%%%%%%%%%%%%%%%%%%

\title{Two-body problem in presence of cosmological constant}

\author{G.S. Bisnovatyi-Kogan}
\address{Space Research Institute, Profsoyusnaya 84/32, Moscow, Russia 117997.\\
National Research Nuclear University MEPhI, Kashira Highway, 31, Moscow, Russia 115409.\\
Moscow Institute of Physics and Technology MIPT,\\
Institutskiy Pereulok, 9, Dolgoprudny, Moscow region, Russia 141701.\\
Department of Physics, University of Rome "La Sapienza", Rome, Italy.}

\author{M. Merafina}
\address{Department of Physics, University of Rome ``La Sapienza", Rome, Italy}

\maketitle

\begin{history}
\received{Day Month Year}
\revised{Day Month Year}
\end{history}

\begin{abstract}
We consider the Kepler two-body problem in presence of the cosmological constant $\Lambda$. Contrary to the classical case, where finite solutions exist for any angular momentum of the system $L$, in presence of $\Lambda$ finite solutions exist only in the interval $0<L< L_{lim}(\Lambda)$. The qualitative picture of the two-body motion is described, and critical parameters of the problem are found. Application are made to the relative motion of the Local Group and Virgo cluster.
\end{abstract}

\keywords{galaxies; two-body problem; cosmological constant.}

\ccode{PACS numbers: 95.36.+x; 98.80.Es; 98.65.At}

%\tableofcontents

%%%%%%%%%%%%%%%%%%%%%%%%%%%%%%%%%%%%%%%%%%%%%%%%%%

%%%%%%%%%%%%%%%%% BODY OF PAPER %%%%%%%%%%%%%%%%%%

\section{Introduction}

It was shown\cite{2001Chernin,2008Chernin} that outer parts of galaxy clusters (GC) may be under strong influence of the dark energy (DE), discovered by observations of SN Ia at redshift $z\le 1$\cite{1998Riess,1999Perlmutter}\,, and in the spectrum of fluctuations of the cosmic microwave background radiation (CMB)\cite{2003Spergel,2004Tegmark}\,. If we consider a relative motion of two rich clusters, than we should deal with sizes, at which the influence of DE taken in the form of the cosmological constant $\Lambda$, is important. Here we consider a simplified problem of the relative motion of two rich clusters, represented by point gravitating masses.

The full mathematical solution of the problem of two-body motion in presence of $\Lambda$, in quasi-Newtonian approximation, was given in Ref.~\refcite{2013Emelyanov}. We analyze the problem from physical point of view, using different mathematical description, and specify the main critical parameters of the two-body system, which have not been calculated in the above cited paper.

\section{Quasi-Newtonian equations of two-body problem with $\Lambda$}

\begin{figure*}
\centerline{\psfig{file=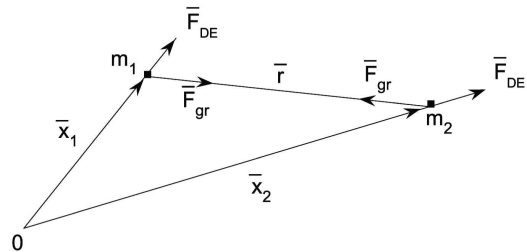,width=7cm}}
\vspace*{0.5cm}
\caption{\small{Arrows indicate direction of the different forces (${\bf F}_{DE}$ and ${\bf F}_{gr}$) on masses $m_1$ and $m_2$ placed in ${\bf x}_1$ and ${\bf x}_2$, respectively.}
\label{fig0}}
\end{figure*}

Here, we reduce the problem to one second order differential equation, similar to the Kepler problem \cite{1969Landau}\,. The form of the contribution of $\Lambda$ in the quasi-Newtonian relativistic approximation was rigorously derived in a previous paper\cite{2014}\,.
Referring to Fig.\,\ref{fig0}, we have for the gravitational force ${\bf F}_{gr}$
$$
{\bf F}_1=\frac{Gm_1m_2}{r^2}\,{\bf\hat{e}}_r\ \ , \ \ {\bf F}_2=\,-\frac{Gm_1m_2}{r^2}\,{\bf\hat{e}}_r\ ;
$$
where ${\bf\hat{e}}_r=({\bf x}_2-{\bf x}_1)/\vert{\bf x}_2-{\bf x}_1\vert$ and $r=\vert{\bf x}_2-{\bf x}_1\vert$.
For the dark energy force ${\bf F}_{DE}$ we have
$$
{\bf F}_1=\frac{\Lambda c^2}{3}\,m_1\,{\bf x}_1\ \ , \ \ {\bf F}_2=\frac{\Lambda c^2}{3}\,m_2\,
{\bf x}_2\ .
$$
Then, the equations of motion for $m_1$ and $m_2$, being ${\bf F}={\bf F}_{gr}+{\bf F}_{DE}$, are given by
$$
m_1\,{\bf\ddot{x}}_1=\frac{Gm_1m_2}{r^2}\,{\bf\hat{e}}_r+\frac{\Lambda c^2}{3}\,m_1\,{\bf x}_1\ ,
$$
$$
m_2\,{\bf\ddot{x}}_2=\,-\frac{Gm_1m_2}{r^2}\,{\bf\hat{e}}_r+\frac{\Lambda c^2}{3}\,m_2\,{\bf x}_2\ .
$$
After some arrangements and introducing the reduced mass $\mu=m_1m_2/(m_1+m_2)$, we obtain
$$
\mu{\bf\ddot{r}}=\,-\frac{Gm_1m_2}{r^2}\,{\bf\hat{e}}_r+\frac{\Lambda c^2}{3}\mu r{\bf\hat{e}}_r\ ,
$$
with ${\bf\ddot{r}}=\,-{\bf\nabla}\Phi$ and $\Phi=\Phi_{gr}+\Phi_{\Lambda}$, being ${\bf F}_{gr}=\,-\mu{\bf\nabla}\Phi_{gr}$ and ${\bf F}_{DE}=\,-\mu{\bf\nabla}\Phi_{\Lambda}$\ .
In this way we reduced the problem to a one body motion, with mass $\mu$ at distance ${\bf r}$ from a central point. Moreover, since
$$
{\bf r}=r{\bf\hat{e}}_r\  ;  \ {\bf\dot{r}}=\dot{r}{\bf\hat{e}}_r+r\dot{\varphi}{\bf\hat{e}}_{\varphi}\  ; \ \ {\bf\ddot{r}}=(\ddot{r}-r\dot{\varphi}^2){\bf\hat{e}}_r+(2\dot{r}\dot{\varphi}+r\ddot{\varphi}){\bf\hat{e}}_{\varphi}\ ,
$$
substituting to equation of motion, we have
$$
(\ddot{r}-r\dot{\varphi}^2){\bf\hat{e}}_r+(2\dot{r}\dot{\varphi}+r\ddot{\varphi}){\bf\hat{e}}_{\varphi}=\frac{1}{\mu}\left[\,-\frac{Gm_1m_2}{r^2}+\frac{\Lambda c^2}{3}\mu r \right]{\bf\hat{e}}_r
$$
and, finally, introducing $M=m_1+m_2$, we obtain for each component
$$
\ddot{r}-r\dot{\varphi}^2=\,-\frac{GM}{r^2}+\frac{\Lambda c^2}{3}r
$$
$$
2\dot{r}\dot{\varphi}+r\ddot{\varphi}=0\ .
$$
The second equation implies that
$$
\dot{L}=\mu r^2 \ddot{\varphi}+2\mu r\dot{r}\dot{\varphi}=\mu r (r\ddot{\varphi}+2\dot{r}\dot{\varphi})=0
$$
and therefore $L=\rm{constant}$. Thus we can write
$$
\dot{\varphi}=\frac{L}{\mu r^2}
$$
and the first equation of motion becomes
$$
{\ddot r} = \,-\frac{GM}{r^2} +\frac{L^2}{\mu^2 r^3}+\frac{\Lambda c^2}{3} r\ .
$$
In analogy with the relation ${\bf\ddot{r}}=\,-{\bf\nabla}\Phi$, where $\Phi=\Phi_{gr}+\Phi_{\Lambda}$, we can impose
$$
\ddot{r}=\,-\frac{d\Phi_{eff}}{dr}\ ,
$$
where the effective potential $\Phi_{eff}$ also includes centrifugal term
$\Phi_c$
$$
\Phi_{eff}=\Phi+\Phi_c=\Phi_{gr}+\Phi_{\Lambda}+\Phi_c\ .
$$
Therefore, we obtain
$$
-\frac{GM}{r^2}=\,-\frac{d\Phi_{gr}}{dr}\ \ ; \ \ \frac{L^2}{\mu^2 r^3}=\,
-\frac{d\Phi_c}{dr}\ \ ; \ \ \frac{\Lambda c^2}{3} r=\,-\frac{d\Phi_{\Lambda}}{dr}\ .
$$
For the gravitational term we have
$$
d\Phi_{gr}=\frac{GM}{r^2}\,dr\ \Rightarrow\ \int_r^\infty d\Phi_{gr}=
\int_r^\infty \frac{GM}{r^2}\,dr
$$
$$
=\,-\frac{GM}{r}\bigg\vert_r^\infty=
\Phi_{gr}(\infty)-\Phi_{gr}(r)=\frac{GM}{r}\ ,
$$
with the condition of normalization $\Phi_{gr}(\infty)=0$ leading to
$$
\Phi_{gr}(r)=\,-\frac{GM}{r}\ .
$$
For the centrifugal term we have
$$
d\Phi_{c}=\,-\frac{L^2}{\mu^2 r^3}\,dr\ \Rightarrow\ \int_r^\infty d\Phi_{c}=\,
-\int_r^\infty \frac{L^2}{\mu^2 r^3}\,dr
$$
$$
=\frac{L^2}{2\mu^2 r^2}\bigg
\vert_r^\infty=\Phi_{c}(\infty)-\Phi_{c}(r)=\, -\frac{L^2}{2\mu^2 r^2}\ ,
$$
with the condition of normalization $\Phi_{c}(\infty)=0$ leading to
$$
\Phi_{c}(r)=\frac{L^2}{2\mu^2 r^2}\ .
$$
For the dark energy term we have
$$
d\Phi_{\Lambda}=\,-\frac{\Lambda c^2}{3}r\,dr\ \Rightarrow\
\int_0^r
d\Phi_{\Lambda}=\,-\int_0^r \frac{\Lambda c^2}{3}r\,dr
$$
$$
=
\,-\frac{\Lambda c^2}{6}r^2\bigg\vert_0^r
=\Phi_{\Lambda}(r)-\Phi_{\Lambda}(0)=\, -\frac{\Lambda c^2}{6}r^2\ ,
$$
with the condition of normalization\cite{2012Merafina} $\Phi_{\Lambda}(0)=0$ leading to
$$
\Phi_{\Lambda}(r)=\, -\frac{\Lambda c^2}{6}r^2\ .
$$

\section{Integrals of motion and stability problem}

The expression of the kinetic energy $T$ is given by
$$
T=\frac{\mu v^2}{2}\ \ {\rm where}\ \
$$
$$
v^2={\bf\dot{r}}\cdot{\bf\dot{r}}=(\dot{r}{\bf\hat{e}}_r+r\dot{\varphi}{\bf\hat{e}}_{\varphi})\cdot (\dot{r}{\bf\hat{e}}_r+r\dot{\varphi}{\bf\hat{e}}_{\varphi})= \dot{r}^2+r^2\dot{\varphi}^2\ ,
$$
then we obtain $T=\mu(\dot{r}^2+r^2\dot{\varphi}^2)/2$, with $\dot{\varphi}=L/\mu r^2$ and, finally, we can write
$$
T=\frac{1}{2}\mu\dot{r}^2+\frac{L^2}{2\mu r^2}\ .
$$
Centrifugal potential is included in the expression of kinetic energy, being
$\mu\Phi_c=L^2/2\mu r^2$. We have
$$
T=\frac{1}{2}\mu\dot{r}^2+\mu\Phi_c
$$
and, therefore, the potential energy is
$$
U=\mu (\Phi_{gr}+\Phi_{\Lambda})=\,-\frac{GM\mu}{r}-\frac{\Lambda c^2}{6}\mu r^2\ ,
$$
while the total energy is given by
$$
E=T+U=\frac{1}{2}\mu\dot{r}^2+\mu\Phi_{eff}\ ,
$$
where
\be
\Phi_{eff}(r)=\,-\frac{GM}{r}+\frac{L^2}{2\mu^2 r^2}-\frac{\Lambda c^2}{6}r^2\ .
 \label{eq2}
\ee
The equation of motion for two bodies $m_1$ and $m_2$ at distance $r$ can be written in the form
\be
 {\ddot r} + \frac{GM}{r^2} -\frac{L^2}{\mu^2 r^3}-\frac{\Lambda c^2}{3} r=0\ ,
 \label{eq1}
 \ee
where $M$ is the total mass given by $M=m_1+m_2$ and $\mu$ is the reduced mass.
The Eq.\,(\ref{eq1}) has a first integral, representing the energy conservation law
$$
\frac{dE}{dt}=\mu{\dot r}{\ddot r}+\mu{\dot r}\frac{d\Phi_{eff}}{dr}=\mu{\dot r}
({\ddot r}+\frac{d\Phi_{eff}}{dr})=0\ ,\ \
$$
$$
{\rm with}\ \ {\ddot r}=\,-
\frac{d\Phi_{eff}}{dr}\ .
$$

\subsection{Circular orbits}

The system with a potential $\Phi_{eff}$ has an equilibrium solution, satisfying the relation

\be
\frac{d \Phi_{eff}}{dr}=\frac{GM}{r^2}-\frac{L^2}{\mu^2 r^3}-\frac{\Lambda c^2}{3}r=0,
 \label{eq3}
 \ee
which corresponds to circular orbits. These are stable if

\be
\frac{d^2 \Phi_{eff}}{dr^2}=-\frac{2GM}{r^3}+\frac{3L^2}{\mu^2 r^4}-\frac{\Lambda c^2}{3}>0 \; .
 \label{eq4}
 \ee
Stationary equilibrium solution in the two-body problem corresponds to a circular orbit, and the solution, corresponding to zero right side in (\ref{eq4}) divide stable and unstable circular
orbits. Multiplying Eq.\,(\ref{eq3}) and (\ref{eq4}) by ($-3r^3/\Lambda c^2$) and ($-3r^4/\Lambda c^2$) respectively, we obtain two equations determining the last stable circular orbit of the system, corresponding to the larger possible radius, in the form

 \be
 r^4-3\frac{GM}{\Lambda c^2}r + 3\frac{L^2}{\mu^2\Lambda c^2}=0,
 \label{eq5}
 \ee
 \be
 r^4+6\frac{GM}{\Lambda c^2}r - 9\frac{L^2}{\mu^2\Lambda c^2}=0.
 \nonumber
 \ee
The system (\ref{eq5}) has three independent parameters: $M,\,\,L,\,\,\Lambda$.   To find the radius of the last stable orbit $r_{stab}$ we may combine these equation in different ways,
 obtaining three different solutions
 \be
 r_{stab,1}=\left(\frac{3GM}{4\Lambda c^2}\right)^{1/3}\, ,
 \nonumber
 \ee
 \be
 r_{stab,2}= \left(\frac{L^2}{\mu^2\Lambda c^2}\right)^{1/4}\, ,
 \label{eq6}
 \ee
 \be
 r_{stab,3}=\frac{4L^2}{3GM\mu^2}\, .
 \nonumber
 \ee
 Substituting each solution from (\ref{eq6}) in one of the equations (\ref{eq5}), we obtain, for a given $M$, the same relation between $\Lambda$ and $L$ for the last stable circular orbit as

 \be
 L_{lim}=\left(\frac{3}{4}\right)^{2/3}\frac{(GM)^{2/3}}{(\Lambda c^2)^{1/6}} \mu, \quad \Lambda_{lim}
 \label{eq7}=\left(\frac{3}{4}\right)^4\frac{(GM)^4 \mu^6}{L^6 c^2} \; .
 \ee
 Note that the same relations follow from each of relations $r_{stab,1}=r_{stab,2}=r_{stab,3}$, what means that we have a unique solution for
 the radius of the last stable orbit $r_{stab}$ in the two-body problem, in presence of $\Lambda$. In the classic Kepler problem the orbits are  always stable.
 Formally, it follows from (\ref{eq6}) that the radius of the last stable orbit $r_{stab}$ goes to infinity when  $\Lambda$ goes to zero.

In presence of $\Lambda$ there is a distance $r_0$ between two bodies at which the gravity is exactly balanced by the antigravity of DE, which is called zero-gravity radius\cite{2008Chernin}\,. The expression for $r_0$ follows from the equilibrium equation (\ref{eq3}) at $L=0$

\be
r_0=\left(\frac{3GM}{\Lambda c^2}\right)^{1/3}=4^{1/3}r_{stab,1}\, .
\label{eq8}
\ee
It follows, that the radius of the last stable orbit is situated inside the zero gravity radius, so that
$r_0=4^{1/3}r_{stab}=1.59\,r_{stab}$. Note, that stable circular orbits exist only at $L<L_{lim}$ at given
$\Lambda$, and at $\Lambda<\Lambda_{lim}$ at given $L$, see (\ref{eq7}). Stable non-circular orbits exist also outside $r_{stab}$, but only inside the zero-gravity radius $r_0$ from (\ref{eq8}), as well as unstable circular orbits\cite{2013Emelyanov}\,.

When the cosmological constant $\Lambda=0$, we have the Keplerian two-body problem in which it exists only a minimum of the potential (stable circular orbit) depending on the value of $L$. The radius of such orbits is defined as

\be
r_k=\frac{L^2}{GM\mu^2} \; .
\label{eq8b}
\ee
%==================================================

\subsection{Stability analysis of circular orbits}
In this section we will study the stability of the possible circular orbits. These orbits are those of which the radius corresponds to minima and maxima of the effective potential.
The first equation (\ref{eq5}) can be written as

\be
r^4-r_0^3r+r_kr_0^3=0 \; ,
\label{9}
\ee
The roots of this equation determine extremes of the effective potential, where circular orbits are situated. The minima of $\Phi_{eff}$ correspond to stable, and the maxima, to unstable circular orbits. Stable non-circular orbits occupy the region inside the zero-gravity radius $r_0$ determined by Eq. (\ref{eq8})\cite{2013Emelyanov}\,.
%Other available orbits in the two body-problem are situated between these extremes.
Because of the complexity of the analytical solution of the Eq. (\ref{9}) and in order to give a widely viewpoint of all the possible circular orbits, here we give the solution of the Eq. (\ref{9}) in terms of $r_0$ and $r_k$ and then, by discriminating two cases:

\be
\rm{(a)} \; \Lambda=const \qquad \rm{and} \qquad \rm{(b)}\; L=const.
\ee
In the first one, we will search the solution as a fraction of the zero-gravity radius $r_0$ and in second one the solution will be give as a fraction of the Keplerian radius $r_k$.
%==================================================
\subsubsection*{(a) {\bf $\Lambda=$const}}

\noindent
In the paper by Emelyanov et al.\cite{2013Emelyanov} it was indicated that the circular orbits situated at $r_{stab}<r<r_0$ are unstable. Here we give a rigorous proof of this statement. We search the solution of the first Eq. (\ref{eq5}), corresponding to the circular orbits, in the form

\be
r_{circ}=\frac{1}{x}\left(\frac{3GM}{\Lambda c^2}\right)^{1/3}=\frac{r_0}{x} \; .
\label{eq9}
\ee
Substituting the equation (\ref{eq9}) into the first equation (\ref{eq5}), we obtain the relations between L, $\Lambda$ and $x$, for circular orbits in the two-body system, in the form

\be
\frac{L^2}{\mu^2 GM}\left(\frac{\Lambda c^2}{3GM}\right)^{1/3}=\frac{x^3-1}{x^4} \; .
\label{eq10}
\ee
It follows from (\ref{eq10}), that circular orbits exist only at $x > 1$.
Here, the Eq. (\ref{9}) can be written as

\be
r_k=r_{circ}\left[1-\left(\frac{r_{circ}}{r_0}\right)^3\right]=r_{circ}\left(1-\frac{1}{x^3}\right)\; ,
\ee
where we use $r_0$ as a fixed value.
The relative velocity of stars on the circular orbit $v_\phi$ is written as

\be
v_\phi^2=\frac {L^2}{r_{circ}^2}=\frac{x^3-1}{x^2}\left(\frac{1}{3}\right)^{1/3}(GM)^{2/3}\mu^{2}(\Lambda c^2)^{1/3}.
\label{eq11}
\ee
At $x\rightarrow 1$ the velocity on the circular orbit goes to zero, the period goes to infinity, so at $x=1$ the equilibrium state of the pair of stars is a static state.
This state corresponds to the zero-gravity radius $r_0$.
As follows from Eqs. (\ref{eq8}) and (\ref{eq9}),

\be
r_{circ}(x=1)=r_0.
\label{eq12}
\ee
To make a conclusion about the stability of the circular orbits, we need to check the sign of the second derivative of the
effective potential $\Phi_{eff}$. We find, using $\Phi_{eff}$ from Eq. (\ref{eq2}) and (\ref{eq10}):

\be
\frac{d^2 \Phi_{eff}}{dr^2}=-\frac{\Lambda c^2}{3}-2\frac{GM}{r^3}+\frac{3L^2}{\mu^2 r^4} = \Lambda c^2 \frac{x^3-4}{3}.
\label{eq13}
\ee
The second derivative is positive only at $x>4^{1/3}$, what corresponds to the orbits inside the last stable circular orbit $r_{circ}<r_{stab}$ from
Eq. (\ref{eq6}). These orbits correspond to minima of the effective potential. The circular orbits in the region $1<x<4^{1/3}$, $r_{stab}<r_{circ}<r_0$ have
a negative value of the
second derivative. They are unstable, and  correspond to maxima of the effective potential.
%This means that for the same value of $L>0$ the effective potential shows at the same time, %minima and maxima until a limit value $L_{lim}$ due to the first equation of (\ref{eq7}).
%==================================================
\subsubsection*{(b) {\bf $L=$const}}

\noindent In this analysis, we search the solution of the first Eq. (\ref{eq5}) as a fraction of the Keplerian radius $r_k$, defined in Eq. (\ref{eq8b}), in the form

\be
r_{circ}=\frac{r_k}{x} \; .
\ee
Once again, substituting this relation in the first equation of (\ref{eq5}) we obtain

\be
\frac{\Lambda c^2}{3GM}\frac{L^6}{\mu^6 (GM)^3}=x^3(1-x) \; .
\ee
From this equation it follows that the circular orbits ex\-ist only for $x<1$.
By keeping constant the value of $L$ and then of the Keplerian radius $r_k$, the Eq. (\ref{9}) can be written as

\be
r_0=\left(\frac{r_{circ}^4}{r_{circ}-r_k}\right)^{1/3} \; .
\ee
To study the stability, it is necessary the analysis of the sign of the second derivative, as in the case (a). It is due to

\be
\frac{d^2 \Phi_{eff}}{dr^2}=\frac{(GM)^4\mu^6}{L^6}x^3(4x-3)>0 \; ,
\ee
which is positive for $x>\frac{3}{4}$. As we said previously, if the request $\frac{d^2 \Phi_{eff}}{dr^2}>0$ is satisfied, the circular orbits are stable. We can conclude that, by keeping constant the value of the angular momentum $L$, all the orbits in which $\frac{3}{4}<x<1$ are stable. This condition corresponds to have radii of the cicular orbits in the range $r_k<r_{circ}<r_{stable}$ (minima of the effective potential). On the other hand, the condition $x<\frac{3}{4}$ is valid for the unstable circular orbits, corresponding to $r_{circ}>r_{stab}$ (maxima of the effective potential).

%=============================================================
\section{Non-dimensional consideration}
In order to illustrate the results obtained above and give a simple physical interpretation, it is convenient to write the dimensionless form of the effective potential. We have to stress that the two different treatment are two faces of the same coin and they are needed to make easier the comprehension of the solutions.
The non-dimensional analysis can be done by using $r_0$ and $r_k$ as scale radii.

%==================================================

\subsection*{Scale radius: $r_0$}
In this case the dimensionless radial coordinate is $\tilde{r}=r/r_0$. The effective potential can be written as

\be
\Phi_{eff}=-\frac{GM}{r_0}\frac{1}{\tilde{r}}+\frac{L^2}{2\mu^2 r_0^2}\frac{1}{\tilde{r}^2}-\frac{\Lambda c^2}{6}r_0^2\tilde{r}^2 \; .
\ee
Substituting $r_0$ we have the dimensionless form of the effective potential, due to

\be
\tilde{\Phi}_{eff}=-\frac{1}{\tilde{r}}+\frac{a}{\tilde{r}^2}-\frac{1}{2}\tilde{r}^2
\ee
where, with account of (\ref{eq8}),(\ref{eq8b}),(\ref{eq9}) we have
\be
\tilde{\Phi}_{eff}=\Phi_{eff} \frac{r_0}{GM}, \qquad a=\frac{1}{2r_0}\frac{L^2}{\mu^2 GM}=\frac{r_k}{2r_0} \; .
\label{50}
\ee
From second equation of (\ref{50}) it follows that for $L \rightarrow 0$ we have $a \rightarrow 0$.
In this case the first relations in Eqs. (\ref{eq6}),(\ref{eq7}) can be written in terms of the parameter $a$, defining this parameter, and the radius of the circular orbit $r_{stab}$ at the boundary of stability, as (see left part of Fig.2)

\be
a_{lim}=\frac{3}{2^{11/3}}\simeq 0.24, \quad \tilde
r_{stab}=\left(\frac{8a_{lim}}{3}\right)=2^{-2/3} \simeq 0.63.
\ee

%==================================================

\subsection*{Scale radius: $r_k$}
To write the dimensionless form of the effective potential $\Phi_{eff}$ we can introduce a dimensionless radial coordinate $\tilde{r}=r/r_k$. Then, the effective potential can be written as

\be
\Phi_{eff}=-\frac{GM}{r_k}\frac{1}{\tilde{r}}+\frac{L^2}{2\mu^2 r_k^2}\frac{1}{\tilde{r}^2}-\frac{\Lambda c^2}{6}r_k^2\tilde{r}^2 \; .
\ee
Substituting $r_k$ from (\ref{eq8b}) we have the dimensionless form of the effective potential, due to

\be
\tilde{\Phi}_{eff}=-\frac{1}{\tilde{r}}+\frac{1}{2\tilde{r}^2}-4a^3\tilde{r}^2 \; ,
\ee
where
\be
\tilde{\Phi}_{eff}=\Phi_{eff} \frac{r_k}{GM} \; .
\label{40}
\ee
From Eqs. (\ref{eq8}),(\ref{50})  it follows that for $\Lambda \rightarrow 0$ there is $r_0 \rightarrow \infty $, $a \rightarrow 0$ and we have the classical Keplerian problem.
The relations from Eqs. (\ref{eq6}),(\ref{eq7}) can be written in this scaling as

\be
a_{lim}=\frac{3}{2^{11/3}}\simeq 0.24, \quad \tilde
r_{stab}=\left(\frac{1}{2^{5/3}a_{lim}}\right)=\frac{4}{3}.
\ee

%===========================================================
\section{Application to the LG and Virgo cluster}

The two-body problem in presence of dark energy can be applied on large cosmological scales (two clusters of galaxies) as well as by considering the motion of a galaxy in a clusters of galaxies.
Let us consider two well known clusters of galaxies, such as Virgo cluster (VC) and the Local Group (LG). In this estimation VC is considered as the only source of gravity, while the gravity of other objects between VC and LG is not important.
We know from observations that the masses of the two clusters are\cite{2015Lee,2008Li,2017Carl}

\be
M_{V}\simeq 10^{15} {\rm M_{\odot}},\qquad {\rm and} \qquad M_{LG}\simeq 10^{12} {\rm M_{\odot}}
\ee
and the distance $d$ between them is estimated in a range $15\div 20$ Mpc\cite{virgo,Solanes}\,.
The total mass of the two bodies is $M\simeq 10^{15} M_{\odot}$ and the reduced mass is $\mu \simeq 10^{12} M_{\odot}$.
From this information we are able to evaluate the zero-gravity radius $r_0$, considering the present value of the cosmological constant $\Lambda$\cite{2008Chernin}

\be
\Lambda=1.3\cdot 10^{-56}\, {\mbox{ cm}}^{-2}, \quad \rho_{\Lambda}=
\frac{\Lambda\,c^2}{8\pi G}
\approx 0.7\cdot 10^{-29}
{\mbox{ g/cm}}^3.
\ee
We have than

\be
r_0=\left(\frac{3GM}{\Lambda c^2}\right)^{1/3}\approx 11\, {\mbox Mpc} \approx 3.3\cdot 10^{25}\, {\mbox{ cm}}.
\ee
So, the Local Group is far outside the zero gravity radius, where DE antigravity is stronger than the gravitational attraction of Virgo cluster.
%We can predict
It is known from observations  that the Local Group is moving toward the Virgo Cluster with velocity around $v_{LG,V}\simeq 250$ km/s\cite{1993Tamman,2010Chernin}\,, on the Hubble expansion background with $v_H\simeq 1000$ km/s \cite{Arp,2000Tammann}\,. With account of it the LG is moving {\bf from} VC at velocity $v_{LV} \sim 750$ km/s. The behaviour of the Local Group in future was predicted in different papers\cite{virloc,virgo1,local} as follows, see also Ref.~\refcite{Loeb}

``{\it Analysis of the dynamics of the Local Supercluster has provided a picture of how the enormous mass of the Virgo Cluster acts gravitationally on the galaxies and galaxy groups around it. As a result, the value of the Virgo Cluster mass and the motion of the Local Group with respect to it (the Virgo-centric flow at the location of our small family of galaxies) seems to indicate that the Virgo Cluster is massive enough to slow down and eventually stop our cosmological recession from it, and then accelerate the Local Group member galaxies toward the central region of the Local, or Virgo Supercluster of Galaxies. Consequently, it appears probable that the Local Group will finally fall and merge into, or be {\bf eaten} by the Virgo Cluster.}"

Basing on the above parameters of Virgo cluster and its distance to LG, the
the second cosmic velocity $v_{2c}$ for VC at the level of LG is equal to

\begin{eqnarray}
v_{2c}=\sqrt{\frac{2GM}{d}}\approx 7.74\cdot 10^7 {\rm cm/s},\quad {\rm at} \quad d=15\,{\rm Mpc}\nonumber\\
\approx 6.7\cdot 10^7 {\rm cm/s},\quad {\rm at} \quad d=20\,{\rm Mpc}.
\end{eqnarray}
It follows, that at $d=15\,$Mpc we have $v_{2c}>v_{LV}$, leading to merger of LG and VC. Estimate the time of merging, using the integral of motion  with radial velocity only, at $\Lambda=0$ and $d=15\,$Mpc, written as

\begin{eqnarray}
\frac{{\dot r}^2}{2} -\frac{GM}{r}=\frac{{v_{r0}}^2}{2} -\frac{GM}{d}
= D =-1.54\cdot 10^{14}\,\rm{cm^2/s^2}.
\label{eqq10a}
\end{eqnarray}
The gravity of VC stops Hubble recession of LG ($\dot r=v_{LG}=0$) at the distance $r_1\approx$ 290 Mpc from VC. After that it starts to approach VC until merging. Elementary calculations give the approximate expression of the time of merging $t_m$, starting from the present moment, as

\be
t_m\approx \frac{2\pi GM}{(2D)^{3/2}}\approx 2\cdot10^{20} \, {\rm s},
\ee
what is much larger than the cosmological time $t_m\approx 5\cdot 10^{17}$ s. This conclusion cannot be taken seriously, because of observational errors in the values of distance $d$ and  mass $M_{VC}$, on which  this number depends very strongly.

At $d=20\,$Mpc we have $v_{2c}<v_{LV}$. Here the gravity of the Virgo cluster produces a considerable decrease in the recessing velocity between LG and VC. The integral of motion is written as

\begin{eqnarray}
\frac{1}{2}{\dot r}^2 -\frac{GM}{r}=\frac{1}{2}{v_{r0}}^2 -\frac{GM}{d}
= D =5.9\cdot 10^{14}\,\rm{cm^2/s^2}.
\label{eqq10}
\end{eqnarray}
The asymptotic value of the recessing velocity between LG and VC is tending to

\be
v_{r1}=\sqrt{2D}\approx 344\,\rm{km/s},
\ee
decreasing from the value of $\sim 400\,\rm{km/s}$, because of gravitational action of VC. This conclusion is valid until the gravity of surrounding matter starts to be important, what happens when Hubble velocity of LG relative to VC, in the flat universe, is becoming less than the $v_{r1}$. Note, that after discovery of dark energy the results of this section have only historical interest.

\subsection{Influence of $\Lambda$}

Let us consider the case when there is no transversal velocity, and LG, with account of the Hubble expansion is shifting away  from the Virgo cluster with the velocity $v_{r0}\approx 750$ km/s, at angular momentum $L=0$. The integral of motion, with account of (\ref{eq2}) at $r=d=20$ Mpc, in presence of dark energy (see also Ref.~\refcite{Loeb}), is written as

\be
\frac{1}{2}{\dot r}^2 -\frac{GM}{r}-\frac{\Lambda c^2}{6} r^2=\frac{1}{2}{v_{r0}}^2 -\frac{GM}{d}-\frac{\Lambda c^2}{6} d^2 = D =-6.4\cdot 10^{15}\,\rm{cm^2/s^2}.
\label{eqq1}
\ee
It follows from (\ref{eqq1}) that the value of the recession velocity between VC and LG is growing monotonously as

\be
v_r^2= \frac{2GM}{r}+\frac{\Lambda c^2}{3} r^2-12.8\cdot 10^{15}\,\rm{cm^2/s^2},
\label{eqq2}
\ee
tending to de Sitter velocity expansion at $t,r\rightarrow\infty$

\be
v_r\rightarrow \sqrt{\frac{\Lambda c^2}{3}}\, r,
\label{eqq3}
\ee
with de Sitter Hubble constant, and event horizon as
\be
H_{dS}=\sqrt{\frac{\Lambda c^2}{3}}\approx 60\,{\rm{\frac{km/s}{Mpc}}},
\quad d_{hor}=\sqrt{\frac{3}{\Lambda}}\approx 5100\,{\rm Mpc}.
\label{eqq4}
\ee
This value is slightly less than the present value $H\approx 70\,\rm{km\,s^{-1}\, Mpc^{-1}}$ \cite{2015Spergel}\,. The present value of $H$ is larger than $H_{dS}$, but a presence of matter implies decreasing of $H$. The relative density of matter is decreasing with time, and asymptotically $H\rightarrow H_{dS}$ at $t\rightarrow\infty$. Note that these conclusions remain the same, if we take $d=15$ Mpc.

\subsection{The case with nonzero transverse velocity }

Consider the case with a finite transversal velocity of LG relative to Virgo cluster, equal to $v_{t0}=d\cdot\dot \varphi=\beta v_{r0}$. In that case the integral of motion $D_1$ from (\ref{eq2}), and angular momentum integral $L=\mu\, v_t\,r$, at $r=d=20$ Mpc, are written as\cite{1969Landau}
\begin{eqnarray}
\frac{1}{2}({\dot r}^2+r^2\dot\varphi^2) -\frac{GM}{r}-\frac{\Lambda c^2}{6} r^2=\frac{1}{2}
{v_{r0}}^2(1+\beta^2) -\frac{GM}{d}-\frac{\Lambda c^2}{6} d^2=D_1 \nonumber\\
=[2.81\cdot 10^{15}(1+\beta^2)-9.24\cdot 10^{15}]\,\rm{cm^2/s^2},
 \label{eqq5}
\end{eqnarray}
\be
\frac{L}{\mu}=v_t\,r=r^2\,\dot\varphi=v_{t0}\,d=\beta\, v_{r0}\,d
=\beta\cdot 4.5\cdot 10^{33}\,{\rm cm^2/s}.
 \label{eqq6}
\ee
It follows from (\ref{eqq5}),(\ref{eqq6}) that the value of the recession velocity between VC and LG  is also growing monotonously as

\be
v_r^2= \frac{2GM}{r}+\frac{\Lambda c^2}{3} r^2-\frac{L^2}{2\mu^2 r^2}
+5.62\cdot 10^{15}(1+\beta^2)-18.48\cdot 10^{15}\,\rm{cm^2/s^2},
\label{eqq7}
\ee
also tending to de Sitter velocity expansion (\ref{eqq3}) at $t,r\rightarrow\infty$. The conclusions of this section remain valid at different $d,\, M_{VC}$, and with account of
the gravity of surrounding galaxies.

\section{Conclusions}

\begin{enumerate}

\item We considered the Keplerian two-body problem in presence of Dark Energy (identified with the cosmological constant $\Lambda$) introduced as a third additional force.

\item In presence of $\Lambda$, differently from the classical two-body problem in which all the permitted circular orbits are stable, the effective potential shows maxima, corresponding to radii of the unstable circular orbits.

\item Three radii are introduced in order to study the stability of the orbits: the zero-gravity radius $r_0$ defined for $L=0$ (radius of unstable "circular" orbits with zero angular velocity), the Keplerian radius $r_k$ defined for $\Lambda=0$ (radius of stable circular orbits) and $r_{stab}$ defined as the radius of the last stable circular orbit, at which maxima and minima of
     the effective potential merge in the inflection point.

\item The stability analysis of the circular orbits has been made by studying the first and the second derivative of the
effective potential. Because of the complexity of the analytic solution, we investigated the requirement for  circular orbits by searching the solution, firstly in terms of $r_0$ (keeping constant $\Lambda$ by varying $L$) and then in terms of $r_k$ (keeping constant $L$ by varying $\Lambda$).

\item For $\Lambda$=const we found that for each value of the angular momentum in the range $0<L<L_{lim}$ stable circular orbits are possible for $r_{circ}<r_{stab}$, whereas unstable circular orbits are possible for $r_{stab}<r_{circ}<r_0$.
In the case in which we keep $L$=const, for each value of the cosmological constant $\Lambda<\Lambda_{lim}$ unstable circular orbit are allowed for $r_0>r_{circ}>r_{stab}$ and stable circular orbits are possible for $r_k<r_{circ}<r_{stab}$.

\item Analysis of the behaviour of LG in the field of Virgo cluster, in presence of DE, shows, that LG is moving away from VC with acceleration, and at larger time its velocity is linearly growing with radius, tending to de  Sitter exponential growth of the separation between LG and VC with time.

\end{enumerate}

\section*{Acknowledgements}

The work of GSBK was partly supported
by the Russian Foundation for Basic Research
Grants no.17-02-00760, 18-02-00619  and the programm of RAS presidium no.28.

\section*{Figures}

\begin{figure*}
\centerline{\psfig{file=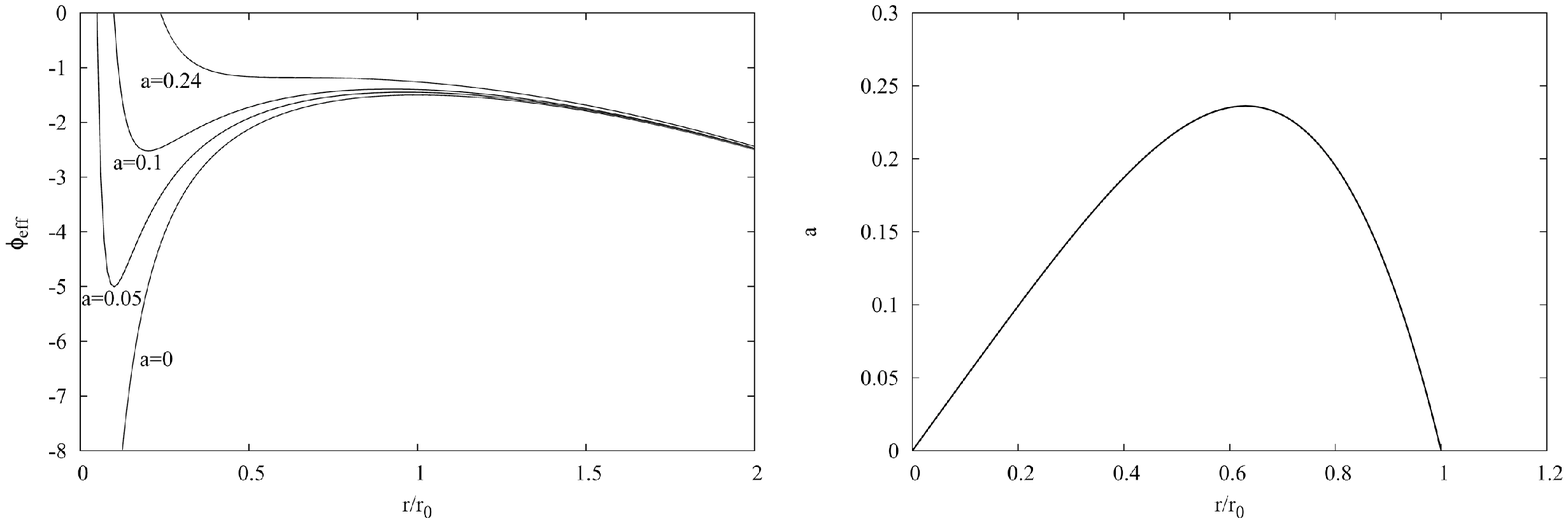,width=12cm}}
\vspace{0.5cm}
%\centering
%\includegraphics[width=12cm]{fig2.eps}
%\vspace{0.8cm}
\caption{\small{\textit{Left}: dimensionless effective potential as a function of $r/r_0$ for different values of the parameter $a=r_k/2r_0$. \textit{Right}: curve of the circular orbits.
The circular orbits exist only at $a\le a_{lim}\simeq 0.24$.
Stable states are represented by minima of the effective potential (left plot), corresponding to increasing part of the curve (right plot). Respectively, maxima of the effective potential, and decreasing part of the curve are related to unstable circular orbits.
%All the increasing values of $r/r_0$ until the maximum (which represents $a_{lim}\simeq 0.24$) are the minima of the %effective potential (stable circular orbits) and the values after $a_{lim}$ are the maxima (unstable circular orbits).
}
\label{fig1}}
\end{figure*}

\begin{figure*}
\centerline{\psfig{file=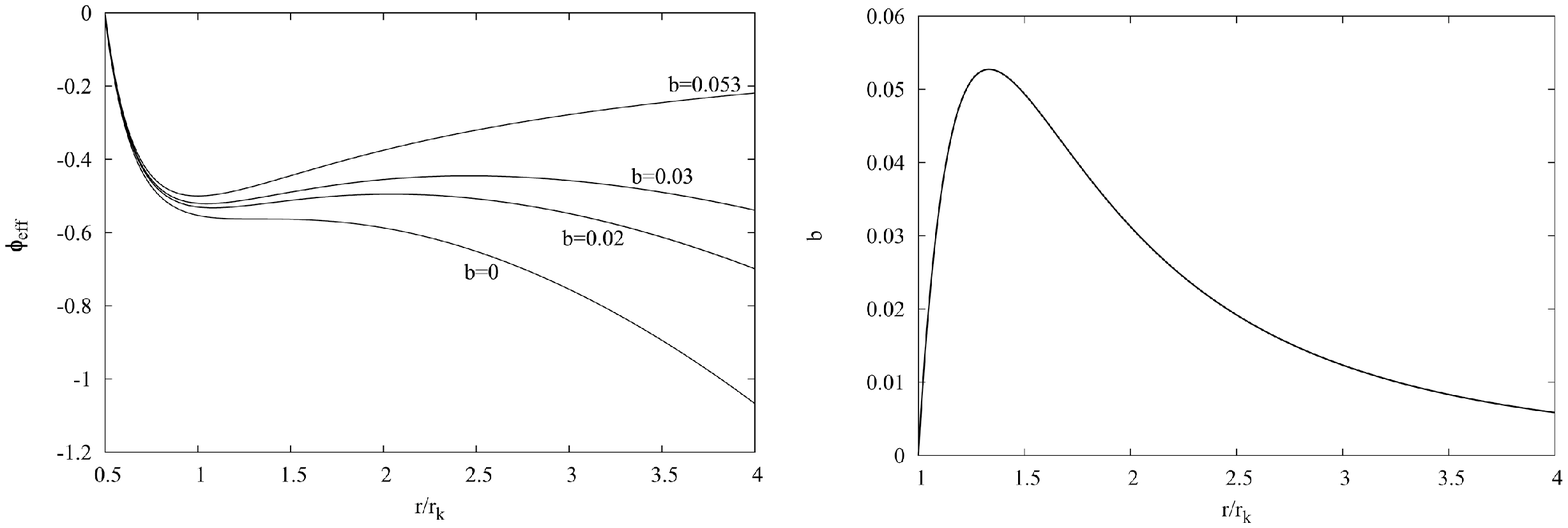,width=12cm}}
\vspace{0.5cm}
%\centering
%\includegraphics[width=12cm]{fig3.eps}
%\vspace{0.8cm}
\caption{\small{\textit{Left}: dimensionless effective potential as a function of $r/r_k$ for different values of the parameter $b=r_k^3/2r_0^3$. \textit{Right}: curve of the circular orbits.
The circular orbits exist only at $b\le b_{lim}\simeq 0.053$.
Stable states are represented by minima of the effective potential (left plot), corresponding to increasing part of the curve (right plot). Respectively, maxima of the effective potential, and decreasing part of the curve are related to unstable circular orbits.
%All the increasing values of $r/r_k$ until the maximum (which represents $b_{lim}\simeq 0.053$) are the minima of the %effective potential (stable circular orbits) and the values after $b_{lim}$  are the maxima (unstable circular orbits).
}
\label{fig2}}
\end{figure*}

\end{document}